\documentclass[conference, a4paper]{IEEEtran}

\usepackage{cite}
\usepackage[pdftex]{graphicx}
\usepackage[cmex10]{amsmath}
\usepackage{amssymb}
\usepackage{algorithmic}
\usepackage{algorithm}
\usepackage{array}
\usepackage{amsthm}
\usepackage[caption=false,font=footnotesize]{subfig}
\usepackage{url}

\begin{document}
\title{Facilitating Machine to Machine (M2M) Communication using GSM Network}

\author{\IEEEauthorblockN{Beena Joy Chirayil}
\IEEEauthorblockA{Intel Comneon GmbH\\Nuremberg, Germany\\
beena.joy.chirayil@intel.com}}

\maketitle

\begin{abstract}

In this paper a method to facilitate M2M communication using existing GSM networks is proposed -- as M2M devices primarily use SMS as their data bearer, the focus is on increasing the number of devices that can use the associated GSM signaling channels at a time. This is achieved by defining a new class of low mobility, static M2M devices which use a modified physical layer control frame structure. The proposal is expected to aid a quick, reliable and cost-effective deployment of M2M devices in the existing GSM networks.
\end{abstract}

\section{Introduction}
Machine to Machine (M2M) Communication is a form of data communication which involves no or only limited human intervention \cite{r3}. M2M communication can be implemented in wireless, wired or hybrid systems. It is also referred to as Machine Type Communication (MTC) in 3rd Generation Partnership Project (3GPP) systems. 

M2M Communication using GSM is promising in terms of cost effectiveness, reliability and mobility. Also as seen from \cite{r3} and \cite{r8}, currently there are no 3GPP standardized solutions for realizing M2M communication using the existing wireless framework.

Short Messaging Service (SMS) is an acknowledged method of textual communication supported by all GSM networks, and it has recently become an important transmission bearer for M2M communication. Its features like encryption, simultaneous transmission with voice-call, and relatively lower cost are added advantages. \cite{r3} covers standardization's study on facilitating M2M communication in 3GPP systems. As seen from \cite{r3}, M2M communication devices use primarily SMS for communication between entities.

The current areas of application of M2M Communication, primarily includes security, surveillance applications, tracking \& tracing, measurement, provisioning and billing of utilities, manufacturing and automation, facility management, remote maintenance. As these applications can considerably increase the number of terminals or devices which are connecting to the network, the Network Operator should also adapt the network entity (example BTS in case of GSM) for handling more number of communication terminals. And as seen from \cite{r3} section  5.2 and \cite{r2}, consideration of handling large number of terminals for M2M communication is still under study by the standardization. 

In this paper we look at a method to handle the increased number of terminals by increasing the number of signaling channels. This is achieved by using a modified control frame structure which multiplexes two users in a manner similar to half-rate speech channels. In section II the modified control frame structure is presented. In section II-A the identification of M2M terminals employing the new control frame structure by a base-station is discussed. In section II-B, a method to signal the control channel sub-allocation to the M2M terminal is proposed. In section II-C, methods to generate the modified control information block of 228 bits along with method for burst-mapping are described.

\section{Modified Control Frame Structure}
As mentioned above, SMS is the primary data bearer for M2M communication and hence hereon we focus on the signaling channels used by SMS in GSM i.e. Stand-Alone Dedicated Control Channel (SDCCH) and Slow Associated Control Channel (SACCH).

\begin{figure}[!h]
\centering
\includegraphics[keepaspectratio=true, width=\columnwidth]{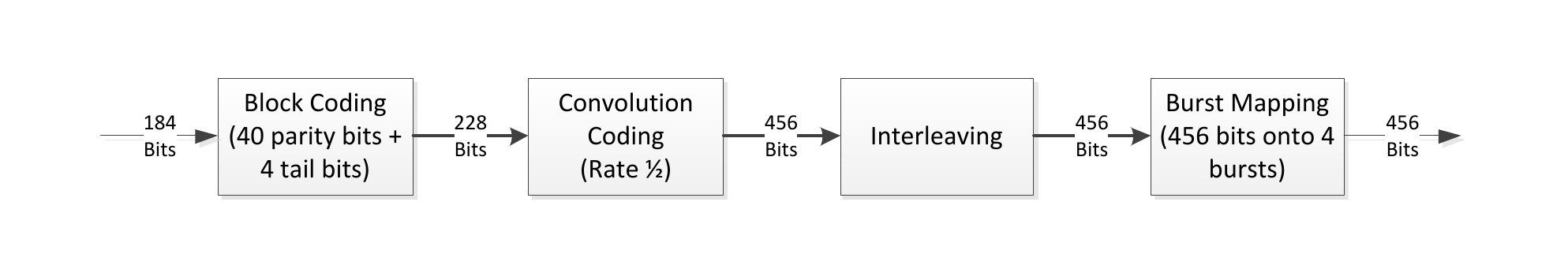}
\caption{Generation of 456 bits control information block}
\label{fig:cib}
\end{figure}

\begin{figure*}[!t]
\centering
\includegraphics[keepaspectratio=true, width=0.95\textwidth]{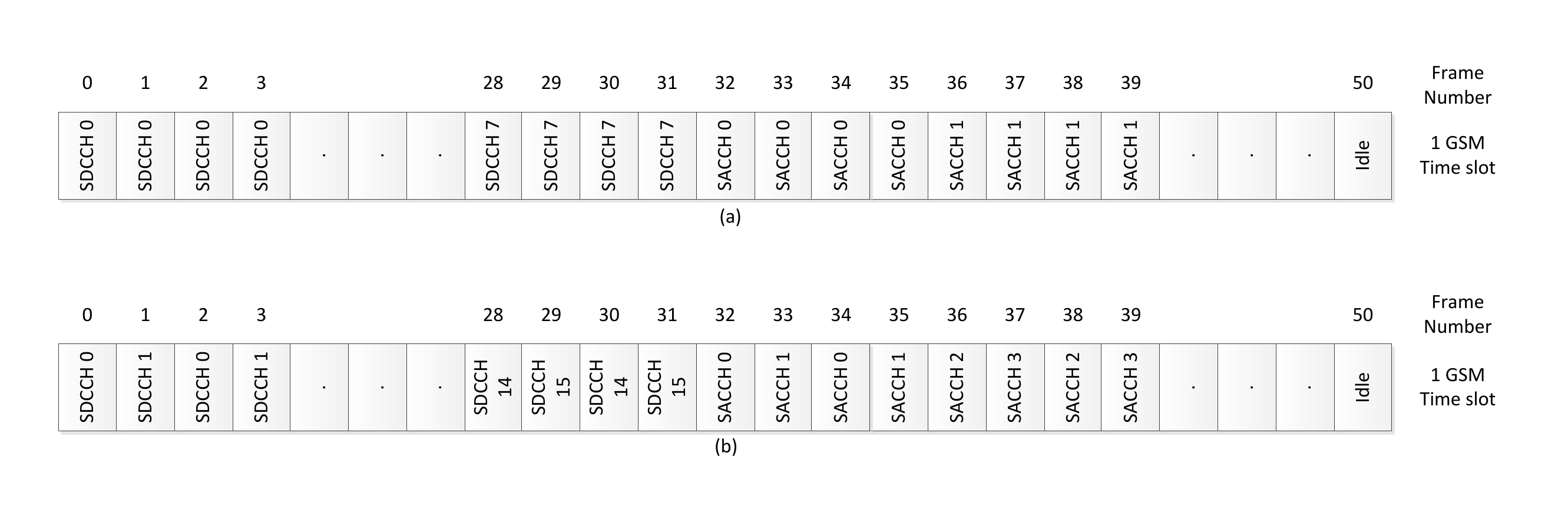}
\caption{GSM 51-multiframe structure for SDCCH/8 configuration (a) original, (b) modified}
\label{fig:multiframe}
\end{figure*}

The SDCCH or SACCH control information message (un-coded) at the physical layer is 184 bits long and is converted into 456 bits control information block after 40 bits block encoding (with 4 tail bits), and rate 1/2 convolution encoding. These 456 bits of the control information block are distributed over four bursts (a control frame) and the four bursts are positioned accordingly in the GSM 51-multiframe structure. Figure \ref{fig:cib} shows the generation of the 456 bits control information block, figure \ref{fig:multiframe}  (a) shows the SDCCH/ SACCH allocation in SDCCH/8 configuration, over a GSM 51-multiframe structure.

\begin{figure*}[!t]
\centering
\includegraphics[keepaspectratio=true, width=0.5\textwidth]{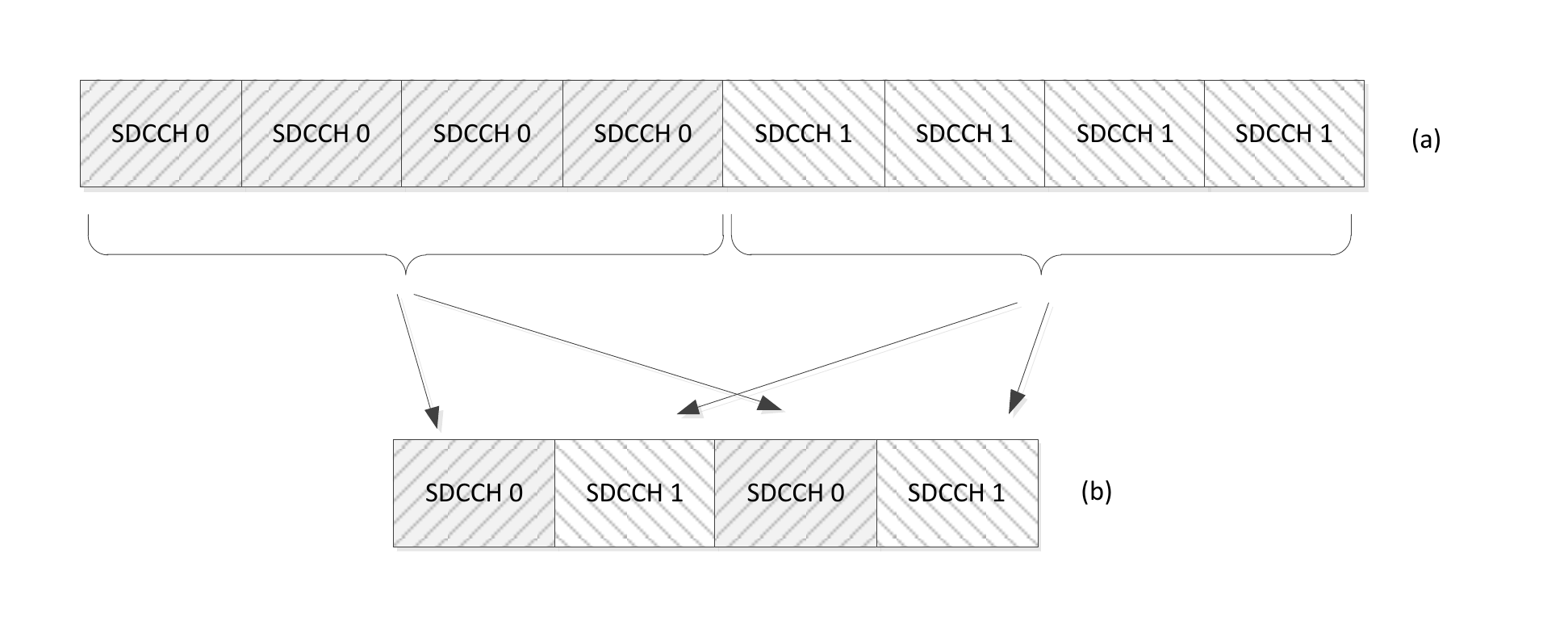}
\caption{Modified mapping of the SDCCH/SACCH control frames of two users onto the four-burst control frame}
\label{fig:bursts}
\end{figure*}

In the modified signaling scheme, for SACCH and SDCCH channels in either SDCCH/8 or SDCCH/4 configurations, the control information block is reduced from 456 bits to 228 bits. These 228 bits are distributed over two bursts, and positioned in GSM 51-multiframe structure as shown in figure \ref{fig:multiframe} (b). Effectively, the original four burst control frame is now shared between two users. As an example, figure \ref{fig:bursts} shows the first user using bursts 0 and 2, and the second user using the bursts 1 and 3. This allocation is similar to the burst-allocation of half-rate speech where the speech channels are multiplexed between two users in a similar manner.

The generation of 228 bit control information block can be done in two ways specific to M2M terminals and is discussed in the later sections.

\subsection{Identification of M2M Terminals}
The GSM base station must be capable of identifying the terminals using the modified signaling scheme. Typically the capability would be signaled to the base-station in the terminal capabilities structure, however, it may also be possible to identify such capable terminals using the International Mobile Subscription Identifier (IMSI) used by the device to register to the network. This is similar to the way an M2M device may be identified using IMSI, as mentioned in \cite{r3}, sections 5.5.1 and 5.5.2.

IMSI is a 15 digit number, consisting of a Mobile Country Code (MCC) of 2 digits, Mobile Network Code (MNC) of 2-3 digits, and Mobile Subscriber Identification Number (MSIN) of 9-10 digits respectively. The M2M terminals using the modified signaling scheme may be identified for e.g. using a unique MNC.

\subsection{Modified Channel Assignment Message}
The channel assignment message is used by GSM network to indicate the physical channel allocations (including control channels) to any mobile station. For the modified signaling scheme, a capable M2M terminal must know its control channel sub-allocation, i.e. whether to use bursts 0, 2 or 1, 3, in addition to the general assignment parameters.

This information may be signaled to the M2M terminal in a modified version of the Immediate Assignment Message where the spare-bit in the 3rd octet of Channel Description IEI may be used to signal the sub-allocation with bit-0 for bursts 0, 2 and bit-1 for bursts 1,3. Refer Table 9.1.18.1 Immediate Assignment Message Content in \cite{r5} for information about the Immediate Assignment Message structure.

\subsection{Modified Control Information Block}
As mentioned above, the modified control information block is 228 bits long. The modified control information block may be generated in two methods.

In the first method, the control information message (un-coded) of 184 bits is used to generate 228 bits instead of 456 bits control information block using puncturing. For static (fixed location), low mobility, low activity M2M terminals (\cite{r3} Section 5.3.2) that operate in good signal conditions, a punctured coding scheme may suffice.

Multiple coding-rate and puncturing-rate schemes may be used to generate the 228 bits from the 184 bits long control information message; a few schemes are mentioned below as examples. An optimal scheme can be selected based on simulations.

\begin{table}[H]
\begin{tabular}{|p{0.05\columnwidth}|p{0.05\columnwidth}|p{0.05\columnwidth}|p{0.2\columnwidth}|p{0.4\columnwidth}|}
\hline
\# & CS & P & Interleaving & Burstmapping \\
\hline
1 & 2/3 & 1/3 & Modified interleaving & 228 bits onto two bursts \\
\hline
2 & 1/2 & 1/2 & Modified interleaving & 228 bits onto two bursts \\
\hline
3 & 1/3 & 2/3 & Modified interleaving & 228 bits onto two bursts \\
\hline
\end{tabular}
\caption{Table of proposed schemes for the first method using the Coding Scheme (CS), Puncturing (P), and Interleaving}
\end{table}

In the second method, the control information message (un-coded) is reduced such that the coded bits are 228 bits long. For M2M devices with low data exchange requirement \cite{r8}, it may be possible to reduce the number of information bits at the data-link layer to fit the message into the modified SDCCH and SACCH control frames using a LAPD$_\textrm{m}$ format \cite{r9} tailored suitably.

For example, with 1/2 convolutional coding, 20 parity, and 4 tail bits, the control information message (un-coded) is 90 bits long and requires a tailored LAPD$_\textrm{m}$ format with 11 octets and 2 filler bits.

\emph{Modified Interleaving and Mapping}:
The block of coded data (228 bits) is interleaved ``block rectangular'' where a new data block starts every 2nd burst and is distributed over 2 bursts. The mapping of 114 bits into a burst remains the same as in \cite{r1} for SDCCH and SACCH channels.

\section{Conclusions}
In this paper a method to facilitate Machine to Machine (M2M) communication in an existing GSM network framework is proposed. The proposed method enables a  GSM  net-\newpage \noindent work  to handle increased flow of signaling that may be generated by M2M devices, thereby enhancing the M2M communication capacity of a GSM network. The proposal will aid in a quick, reliable and cost-effective method for deploying low mobility, static M2M devices in existing GSM networks.

From the 3GPP standardization perspective, changes will be required in \cite{r1}, \cite{r5} and \cite{r9} to support the modified control frame structure, channel assignment message, and control information block generation respectively.

The two methods to generate the modified control information message block provide an implementation tradeoff where the first method retains the data-link layer compatibility but requires good channel conditions, and the second method retains the performance of the existing control channel but requires data-link layer tailoring and is suitable for M2M devices with low data exchange. Further studies are required to evaluate the various coding/puncturing schemes and adaptation of LAPD$_\textrm{m}$ formats.

\end{document}